\documentclass{aa}

\usepackage{graphicx}

\usepackage{txfonts}

\begin{document}

   \title{Cycle dependence of the longitudinal-latitudinal \\ sunspot motion correlations}

   \subtitle{}

   \author{J. Murak\"ozy \and A. Ludm\'any}

   \offprints{J. Murak\"ozy}

   \institute{Heliophysical Observatory of the Hungarian Academy of Sciences
		H-4010 Debrecen P.O.Box 30. Hungary\\
              \email{murakozyj@puma.unideb.hu, ludmany@tigris.unideb.hu}
              }

   \date{}
 
  \abstract
     {}
{It is well known that the azimuthal and meridional shifts of sunspots are correlated and that the correlation exhibits a latitudinal distribution, which is expected due to the Coriolis effect. We study the temporal behaviour of this latitudinal distribution.}
{We analyze the daily positions of sunspot groups, provided by the Debrecen Photoheliographic Data and the Greenwich Photoheliographic Results and correlation values, which were mapped in $ 5^{o} $ latitudinal bins. The latitudinal distributions were examined for each year.} 
{We derive a sunspot-motion correlation that exhibits a Coriolis-type latitudinal distribution on long timescales, which are typical for the yearly distributions; at cycle maximum, however, unexpected distortions can occur.}
{The causes of the weakening of the Coriolis-pattern remain unclear. Possible relations of the phenomenon to the Gnevyshev-gap, the polarity reversal of the main magnetic field, and some mid-period fluctuations are discussed.}


   \authorrunning{Murak\"ozy\&Ludm\'any}
   \titlerunning{Sunspot motion correlations}

   \maketitle

\section{Introduction}

Primarily in rotation analyses, sunspots have long been used as tracers of solar-surface streams. The covariance of latitudinal and longitudinal motions were first investigated by Ward (1965), using sunspot data from the Greenwich Photoheliographic Results (GPR) for the years 1935-1944. He calculated the $\langle v_{\theta} v_{\phi} \rangle$ covariances and reported positive and negative values for the northern and southern solar hemispheres, respectively. This implies equatorward shifts that are produced by positive longitudinal velocities $(v_{\phi})$, because $\theta$ is defined to be the polar angle of the considered feature as measured from the north pole. Based on Sac Peak data, this result was confirmed by Coffey and Gilman (\cite{coffey}), who also published a plot demonstrating the latitudinal growth of covariances with no distinction between the northern and southern hemispheres. Gilman and Howard (\cite{gilman}) analysed Mount Wilson sunspot data for a 62 year period and reported that covariances derived for individual sunspots are smaller (by about 60$\%$) than those derived for sunspot groups. Howard (\cite{howard}) and Pulkkinen and Tuominen (\cite{pulkkinen}) reported almost linear latitudinal variation of covariance for $< 40^{o}$ latitudes. Nesme-Ribes et al. (\cite{nesme-ribes}) measured no significant covariance by using Meudon sunspot measurements for a period of eight years (1977-1984), which contradicts the results of all other studies.

Several authors have used other tracers, such as chromospheric features observed in CaII lines. Belvedere et al. (\cite{belvedere}) argued that faculae observed in the K line are more reliable tracers than sunspots, although their positions are somewhat more ambiguous; they analysed the 1967-70 Catania observations and detected similar behaviour to that reported by Howard (\cite{howard}) and Pulkkinen and Tuominen (\cite{pulkkinen}) . CaII observations were used by Schr\"oter and W\"ohl (\cite{schroter}): they traced the motion of bright mottles and found that their circulation pattern displayed giant cell motions that could be described by $\langle v_{\theta} v_{\phi} \rangle$ covariances. In the above works, the $\langle v_{\theta} v_{\phi} \rangle$ values are approximately $10^{3} m^{2}/s^{2}$,
but the values derived from CaII data are higher than those measured from sunspot data. Vr\v{s}nak et al. (\cite{vrsnak}) detected covariances even in the corona but only for young point-like structures, indicating their anchorage in deeper layers.

Most of the papers cited above measured azimuthal, meridional covariances or correlations using more general terms called the $\langle v_{i} v_{j} \rangle$ turbulent-velocity covariances, also referred to as Reynolds stresses, where $v_{i}$ and $v_{j}$ are the orthogonal velocity components. It is assumed that the sources of Reynolds stresses are turbulent, giant convection cells that interact with the solar rotation by means of the Coriolis force. The Reynolds stresses were held mostly responsible for equatorward momentum transport (Ward (\cite{ward}), R\"udiger et al. (\cite{ruediger})), i.e. for the maintenance of differential rotation. Some authors estimated the equatorward momentum flux using covariance data (Pulkkinen and Tuominen, \cite{pulkkinen}, Patern\`o et al., \cite{paterno}). D'Silva and Howard (\cite{d'silva}) presented an alternative approach and argues that Reynolds stresses are not indispensable in explaining the covariance values: sunspots affected by the Coriolis force may produce similar result, without any turbulent convection pattern. 

Spatial- and temporal-feature properties, such as long-term latitudinal distributions, have been reported, with differences being apparent between hemispheres, and, in some cases, a size dependence on sunspot evolutionary phase being found. We study the cycle dependence of longitudinal and latitudinal motions, where we intend to model temporal behaviour, i.e. its variation during the activity cycle.

\section{Observational material and method of analysis}

We adopt observational data from the Debrecen Photoheliographic Data, the DPD, Gy\H ori et al. (\cite{dpd}), which covers 13 years from 1986 until 1998, and practically the entire solar cycle 22. This catalogue contains position and area data of all observable sunspots on a daily basis and data on sunspot groups, that is the total area of spots in the group and the position of their centre of weight. It is the most suitable source for sunspot studies in cycle 22.

\begin{figure}
   \centering
   \includegraphics [width=8cm] {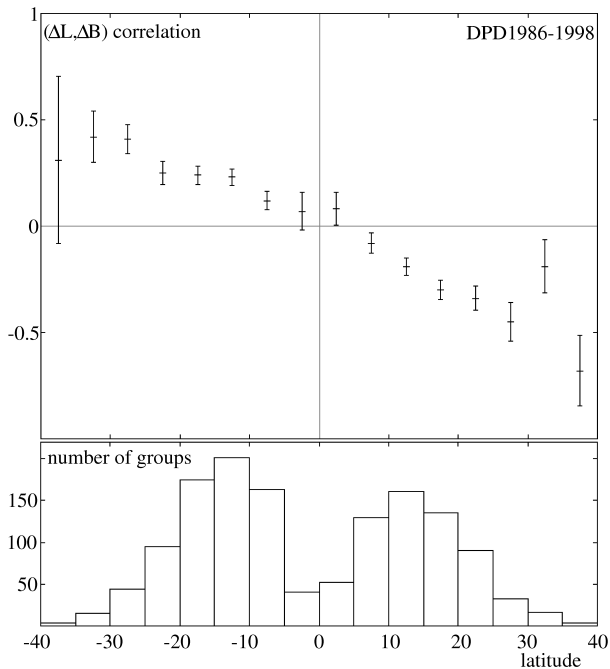}
   \caption{Upper panel: latitudinal distribution of sunspot-motion correlation 
            for the years 1986-1998, lower panel: numbers of sunspot groups within $5^{o}$ latitudinal stripes.}
              \label{8456fig1}
    \end{figure}

The present analysis uses sunspot-group data that may cause a reliability problem, according to Gilman and Howard (\cite{gilman}). These authors argue (following the unpublished criticism of Leighton) that the longitudinal and latitudinal motions of a sunspot group may be produced by morphological features of the groups; in particular, the forward motion of the leading part is usually faster than the receding motion of the following part and this, combined with the well-known tilt of the group axes, may cause an intrinsic equatorward shift, which is not a direct consequence of the Coriolis effect. Using groups as tracers could therefore modify (enlarge) the covariance values. However, the use of single sunspots can modify our analysis results because of the intrinsic (non-Coriolis) sunspot-group rotation. Furthermore, it appears impossible to isolate a Coriolis-effect component in the correlations.

To avoid the impact of the effect mentioned by Leighton, some selection criteria were used. A morphological shift is typical of the growing phase of sunspot-group development and is almost insignificant about the maximum phase, when a group reaches its maximal area and extension: for a while spot emergence and disappearance are insignificant  and the group can be considered an individual entity. In the first approach, we considered only active regions that exhibited maximal area on the visible solar disc and data for a maximum of three days prior to and after the maximal area, depending on observability. This selection should not affect our results because our aim is not to determine the absolute value of the correlations but instead their temporal variation.

Correlations between longitudinal an latitudinal motions were  computed using the formula:

\begin{equation}
      r = \frac{\sum (\Delta L_{i} - \overline {\Delta L} )(\Delta B_{i} - \overline {\Delta B}) }{\sqrt { \sum (\Delta L_{i} - \overline {\Delta L} )^2 } \sqrt { \sum (\Delta B_{i} - \overline {\Delta B} )^2 } }
\,,
   \end{equation}

This formula illustrates the correlation between the $\Delta L_{i}$ and $\Delta B_{i}$ values, the diurnal longitudinal and latitudinal shifts of a given sunspot group, respectively, for the period about which it exhibits its maximal area. The diurnal shifts are differences between the daily positions of sunspot groups taken from the DPD catalogue. Since the observations were completed at different moments of the days the diurnal shifts are normalised to temporal differences of 24 hours. The  $\langle\Delta L,\Delta B\rangle$ covariance has been used for this type of analysis, which is the numerator of the above formula; its dimension is velocity squared and it can be considered to be a measure of the Reynolds stress. The covariance is a suitable tool for studying the magnitude of the effect for the aforementioned theoretical reasons, for studying the temporal behaviour of the effect, however, a normalised quantity, the correlation coefficient, appears to be more informative. 

For similar reasons no correction was made for differential rotation. On the one hand, the rotation profile exerts similar influence during the cycle and therefore does not modify the temporal profile of the correlations. On the other hand, the differential rotation profile varies with depth and it is not obvious which depth should be applied. In any case, it appears informative not to burden the results with ambiguous modifications but to follow a normalised parameter with respect to a steady, rotating frame.

\section{Measurements of cyclic variation}

By using the above procedure, we attempt to find the curve of latitudinal dependence of the $(\Delta L, \Delta B)$ correlations. The first step provides the curve obtained over the entire solar cycle. In the second step, individual curves are plotted on a yearly basis to follow any eventual connection with the cycle phase. If some deviations are obtained from the patterns expected on the basis of the Coriolis effect, then these may be signatures of the impact of a changing velocity field.

A correlation coefficient was computed by using the above formula for each selected sunspot group (by following it from the first selected day to the last one) in each $5^{o}$ wide latitudinal stripe and the derived values were averaged within the stripes in both hemispheres. Figure~\ref{8456fig1} shows the latitudinal distribution of the $(\Delta L, \Delta B)$ correlation coefficient for the years 1986-1998, along with the numbers of considered groups within the $5^{o}$ latitudinal stripes. 

\begin{figure}[h]
   \centering
   \includegraphics [width=8cm]{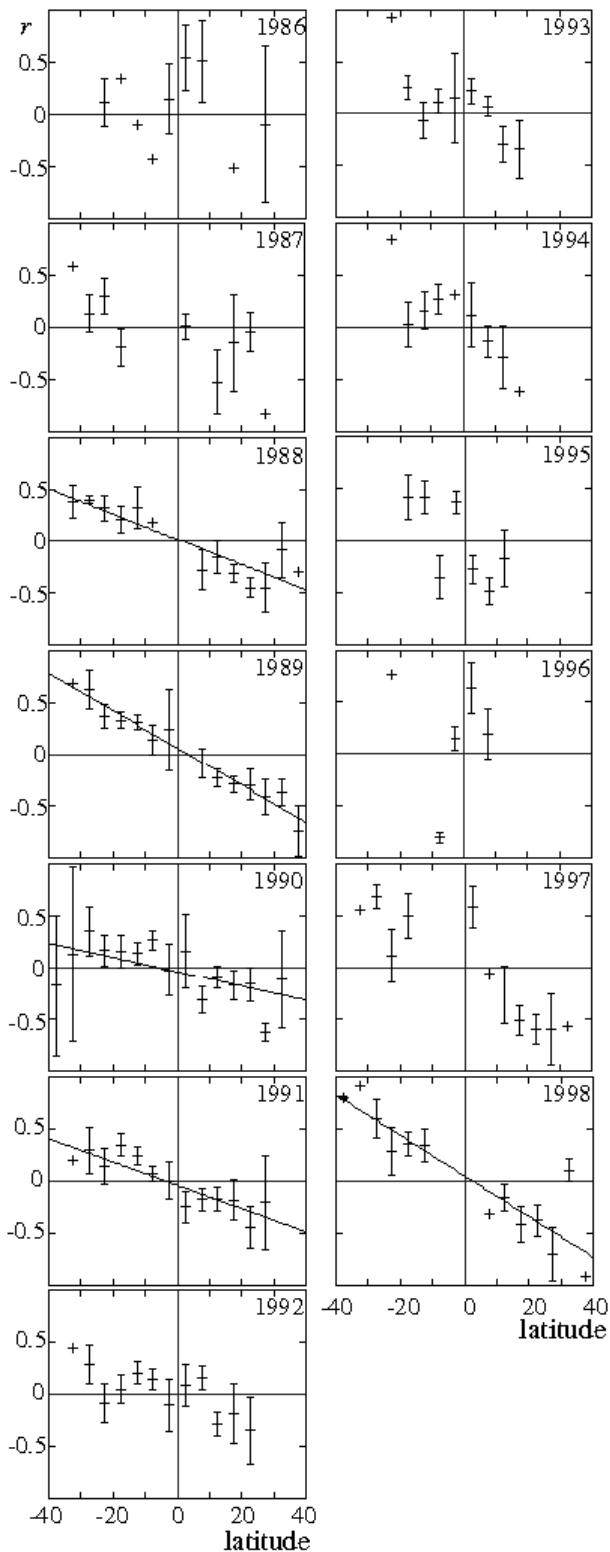}
   \caption{Latitudinal distributions of the $(\Delta L, \Delta B)$ correlations on a yearly basis in the years1986-1998.}
              \label{8456fig2}
    \end{figure}

\begin{figure}[h]
   \centering
   \includegraphics [width=6cm]{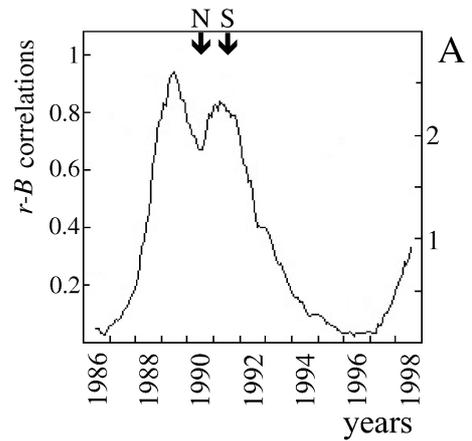}
   \caption{Shape of cycle 22 by smoothed, monthly mean sunspot-area data (from DPD); {\it A} denotes the area in 1000 MSH. The times of magnetic polarity reversals at the northern and southern poles are indicated by arrows.}
              \label{8456fig3}
    \end{figure}

The derived latitudinal distribution is similar to that published by Pulkkinen and Tuominen (\cite{pulkkinen}) (if one takes into account the differences in coordinate definitions) and Latushko (\cite{latushko}). The most interesting feature can be seen in the temporal behaviour. Figure~\ref{8456fig2} shows the plots of the latitudinal distributions of correlations for each of the 13 years separately. The phase of cycle 22 can be traced by comparing the panels of each year with the cycle shape plotted in Fig.~\ref{8456fig3}.

It is conspicuous that the most unanimous monotone latitudinal distribution is found in the maximum year, 1989 and one year before, 1988. The years of increasing activity show a similar trend, with larger scatter because of small-number statistics apart from the years at the end of the declining phase, which exhibit stochastic patterns. The distribution can be observed again after minimum, when the spots of cycle 23 appear at high latitudes (1997 and 1998). These properties may not be too surprising: we indeed expect a disappearing effect close to the equator, where the Coriolis force is zero; this decrease at cycle decay was also found by Gilman and Howard (\cite{gilman}) and Balthasar et al. (\cite{balthasar}).

There is, however, a feature that may deserve attention. As in the entire cycle (Fig.~\ref{8456fig1}), the {\it r} correlation values exhibit a unanimous decreasing trend in the plots of 1988-89, in other words, the correlation decreases with increasing latitude, as expected, and can be regarded as a standard Coriolis pattern. However, this trend appears to be weakened in 1990 and strengthens again in 1991. This cannot be the result of small-number statistics because the number of sunspots is higher than in 1988 and only slightly smaller than in 1989. Moreover, the activity belt occupied by sunspots is at its widest about maximum; we would therefore expect the opposite case of the most unambiguous monotone decreasing trend in 1989-91. Its absence in 1990 may be indicative of more complicated behaviour.

The shape of the cycle (Fig.~\ref{8456fig3}) exhibits the well-known 'Gnevyshev gap', a local depression (in 1990) between the two highest values (1989 and 1991) of the cycle. This feature was first detected by Gnevyshev (\cite{gnevyshev}) and analysed subsequently by several authors (Bazilevskaya et al., \cite{bazilevskaya}; Storini et al.,  \cite{storini}). It would be invaluable to identify any feature related to the 'Gnevyshev gap'. The latitudinal redistribution of shift-correlations in the gap appears to hold a key in understanding the process behind the behaviour.

\begin{figure}
   \centering
   \includegraphics [width=8cm]{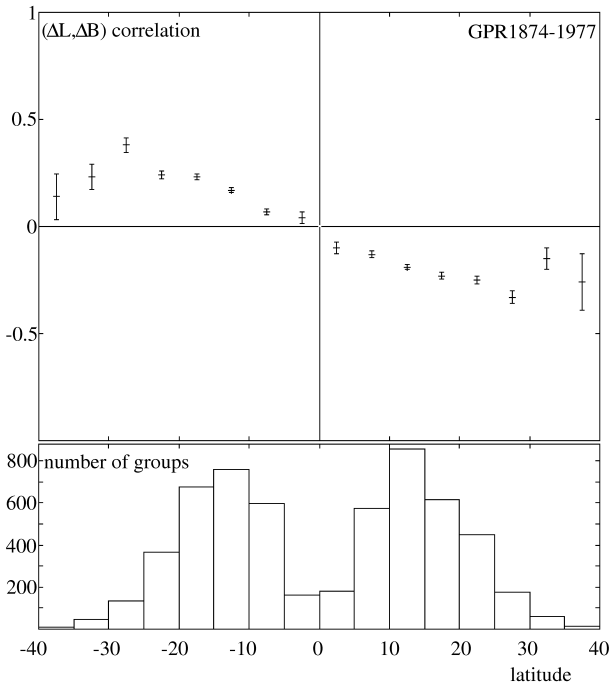}
   \caption{Upper panel: latitudinal distribution of sunspot motion correlation based on GPR data for the years 1874-1977, 
            lower panel: numbers of sunspot groups within the $5^{o}$ latitudinal stripes.}
              \label{8456fig4}
    \end{figure}

For this reason the feature was checked on the Greenwich Photoheliographic Results (GPR, Royal Observatory \cite{gpr}), the classic sunspot catalogue. This step was completed by Pulkkinen and Tuominen (\cite{pulkkinen}) but to provide comparability with results based on DPD, similar selection criteria were used as in the case of DPD data: each sunspot group was considered over an interval containing the day of its maximal area and at least three days before and after that day. Figure~\ref{8456fig4} shows the latitudinal correlation distribution for the entire 1874-1977 period computed in a similar way to in the case of DPD in Fig.~\ref{8456fig1}, along with a histogram showing the numbers of considered sunspot groups in each $5^{o}$ latitudinal bin. The distribution is similar to that in Fig.~\ref{8456fig1}. Comparing the statistics of the two materials we found that, in the GPR, 5683 groups satisfied the selection criteria in 104 years, whereas, in the DPD, 1357 groups were selected over 13 years.

Figure~\ref{8456fig4} has a similar distribution to that of Fig.~\ref{8456fig1} and also Fig. 4 of Pulkkinen and Tuominen (\cite{pulkkinen}) by taking into account the difference in sign conventions. The first five cycles in the years 1874-1977 are much weaker than cycle 22, covered by the DPD: these cycles have far fewer sunspots, so the behaviour about maxima is uncharacteristic and  stochastic due to the small-number statistics. Therefore, cycles 18 and 19 were chosen for monitoring the distributions about the maxima, because they appear to be sufficiently high and exhibit the Gnevyshev-gap, although weakly in cycle 19. Figure~\ref{8456fig6} shows the latitudinal correlation distribution for the years 1946-50 (cycle 18) and 1956-61 (cycle 19), along with plots of the intervals about maxima with the ({\it r,B}) correlations.

\begin{figure}
   \centering
   \includegraphics [width=8cm]{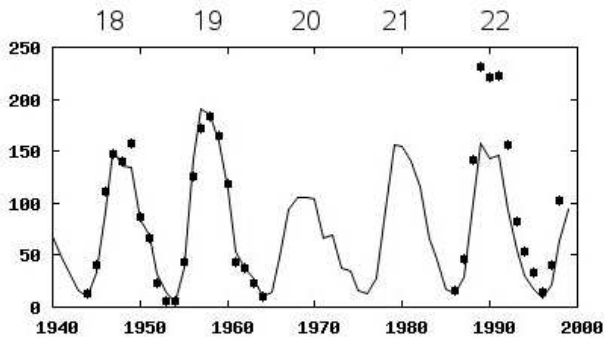}
   \caption{Annual mean sunspot numbers (continuous line) and the annual numbers of applied sunspot groups (dots) for the cycles 18-22.}
              \label{8456fig5}
    \end{figure}

\begin{figure}
   \centering
   \includegraphics [width=8.6cm] {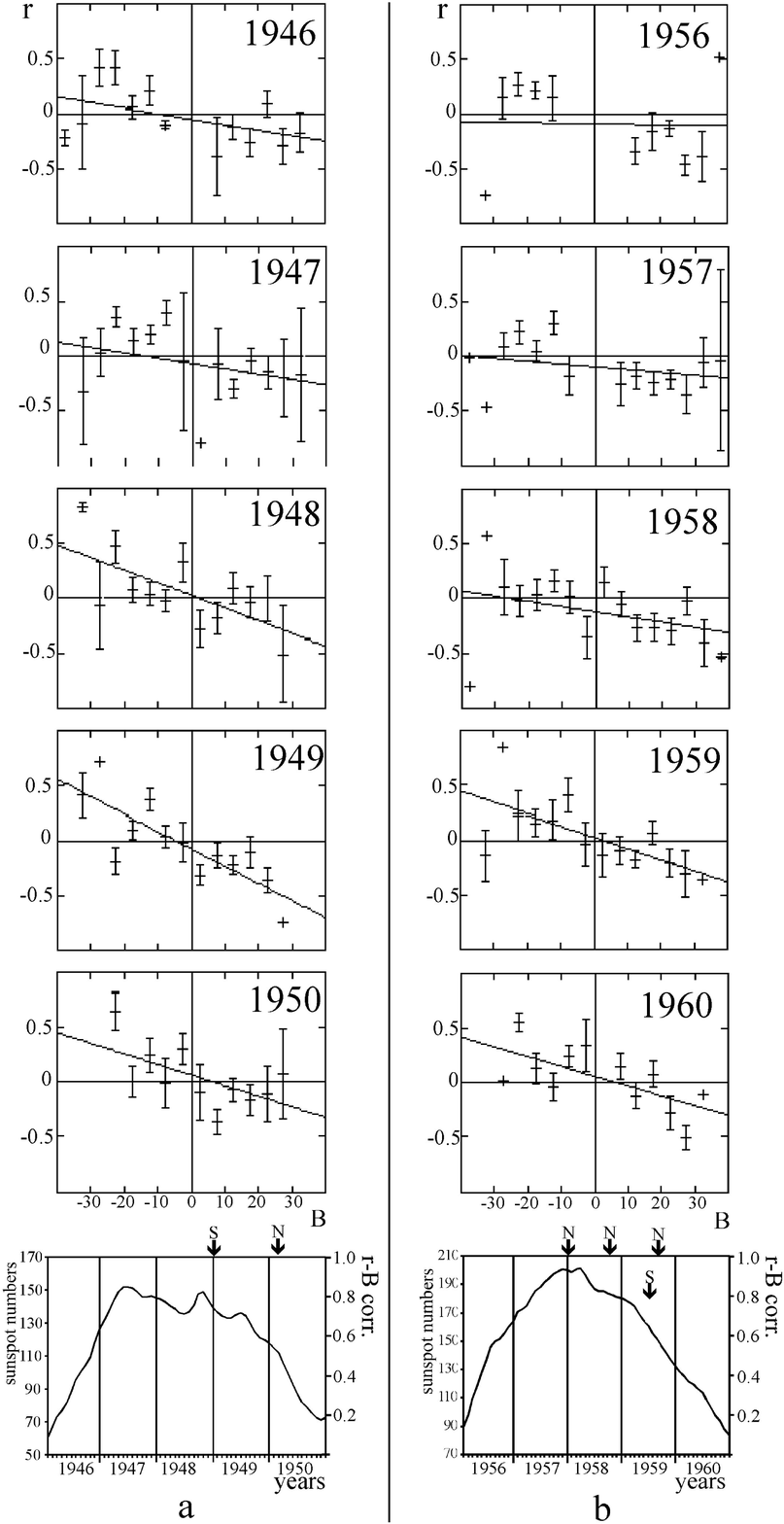}
   \caption{Latitudinal $(\Delta L, \Delta B)$ correlation distribution in the years 1946-1950 (cycle 18, column a.) and 1950-1960 (cycle 19, column b.) for GPR data. The lowest panels show the smoothed monthly sunspot numbers in these years and the times of  magnetic polarity reversals at the northern and southern poles. }
             \label{8456fig6}
    \end{figure}

In cycle 18, the year of rising activity, 1946, and the year of first activity maximum, 1947, both have no standard distribution; the year of declining activity, 1948, and the first year of decay, 1949, do however show a standard distribution. In cycle 19, the years of rising activity 1956-57 do not exhibit a standard distribution, but the year 1958, corresponding to a weak depression close to a maximum, and 1959 do both exhibit a standard distribution. In spite of the remarkable similarity between the cycle profile and the ({\it r,B}) correlations (Fig.~\ref{8456fig3}) around maximum, the data taken from the GPR do not confirm the conjecture that the Gnevyshev gap may cause a drop in the steepness of the latitudinal correlation distribution. 

We note that data for cycles 18/19 and cycle 22 are produced using different procedures and tools: the DPD procedure is highly automated, containing far fewer arbitrary elements than the manual method of GPR. It is interesting to compare the datasets of GPR and DPD in Fig.~\ref{8456fig5}, where the annual mean sunspot numbers and the annual numbers of applied sunspot groups are plotted for cycles 18-22. The mean ratio of the applied groups and the sunspot numbers is 1.12 for the examined GPR period (1874-1977) and 1.59 for the DPD period (cycle 22). It appears that the DPD is a more detailed catalogue. The data statistics are sufficiently high in the aforementioned cycles, however, and any differences should not be automatically attributed to the different methodologies adopted in creating the GPR and DPD catalogues.

\section{Discussion of variations about maxima}

As mentioned in the Introduction, the sunspot-motion covariances were originally considered to be signatures of the Reynolds stresses and their values were used to ascertain the equatorward momentum transport. This approach was criticised by some authors. D'Silva and Howard (\cite{d'silva}) argued that the sunspots do not require turbulent motion of the ambient gas to produce the obtained covariances: they can simply move according to the Coriolis effect. Instead of a momentum-transport analysis, we have focused on temporal variations and have used correlation values instead of covariances. The separation of days about the maximum area of sunspot groups means that the role of emergence (radial motion of the field) is minimized in the Coriolis-turns.

The distributions averaged over the long intervals depicted in Figs.~\ref{8456fig1} and ~\ref{8456fig4} exhibit a pattern expected on account of the Coriolis effect. As for the yearly distributions, the expectations are based on Sp\"orer's law: when active regions are at higher latitudes, the Coriolis effect should be more pronounced than at the end of the cycle decay, close to the equator. This last feature is apparent in Fig.~\ref{8456fig2} (panel of the year 1996). We expect, however, that in the years of activity maximum, when the active regions are spread over the widest latitudinal belts, the distribution of shift-correlations is the most comparable to the distribution averaged over the entire cycle (Fig.~\ref{8456fig1}); this is not however the case. In certain years at maximum, the regular Coriolis pattern may weaken or almost disappear. 

It should be noted that cyclic variations in the $(\Delta L, \Delta B)$ correlation cannot be produced by any cyclic variations in the differential rotation. As Eq. 1 shows, the {\it r} parameter can provide only a relation between the diurnal variations of longitudinal/latitudinal positions and is independent of the magnitude of their values. At a certain latitude, the angular velocity is constant during the period in which the correlation coefficient for a  sunspot is computed (max. 11 days) and the actual value of the angular velocity has no impact on the correlation coefficient. On the other hand, the differential rotation does not exhibit abrupt changes from one year to the next, which is the most interesting feature of correlation distribution. Differences between odd-even cycles cannot play a role, since data from cycles 18 and 22 are different in this respect so the effect appears to be independent of magnetic polarity conditions.

To interpret the fluctuation, three explanations appear to be worth examining. The first idea concerns the role of the Gnevyshev gap (\cite{gnevyshev}). In cycle 22, the drop in curve steepness coincides with the Gnevyshev gap (Fig.~\ref{8456fig2}, year 1990), although, in cycles 18 and 19 (Fig.~\ref{8456fig6}) it does not. In cycle 22, data for the two years prior to the first maximum (1988 and 1989) exhibit the most pronounced example of a standard pattern and the Gnevyshev gap (1990) does not; in cycles 18 and 19 however, the pattern is weak in the years prior to the maxima and strengthens in or after the gap. The Gnevyshev gap is therefore not the cause of this weakening.

The second possible idea concerns the role of the polarity reversal in the main magnetic dipole field. To test this hypothesis, the dates of the northern and southern polarity reversals were indicated in the figures of activity curves reported by Makarov and Makarova (\cite{makarov}) (see Fig.~\ref{8456fig3}, and the last panels of Fig.~\ref{8456fig6}). In all three cases the reversals occured at or after the secondary maxima of the cycles (in cycle 19 three northern reversals were detected, but the final situation had been established at the start of decay). The order of events implies that the polarity reversals cannot play role in either the formation of the Gnevyshev gap or the fluctuation of the steepness in the Coriolis distribution. 

A third possible interpretation is based on the possible interplay between the 11-year cycle and some kind of quasi-biennial fluctuation, which was proposed to explain the Gnevyshev gap by Bazilevskaya et al. (\cite{bazilevskaya}). A wide variety of such fluctuations are reported from tachoclyne zone to cosmic rays, but any relations or interconnections between them remain unclear. Their periods, for example, are quite different. Mursula et al. (\cite{mursula}) used the name of mid-term fluctuations to describe fluctuations of periods shorter than 2 years, whereas Ivanov et al. (\cite{ivanov}) defined quasi-biennial and quasi-triennial fluctuations. To interpret the variations in the correlation distribution, in terms of mid-period fluctuations, a relevant domain should be found that exhibits these kinds of fluctuations and may be able to exert an impact on the velocity correlations. 

A possible candidate to influence these correlations may be an interplay between the radial shear oscillation at the tachoclyne zone (Howe and Christensen-Dalsgaard, \cite{howe}) and the giant cells. The giant cells were found in simulations by Gilman and Glatzmaier (\cite{gilman}). Its observational detection, however, remained difficult and results are not yet conclusive (Baranyi and Ludm\'any, \cite{baranyi};  Beck et al.\cite{beck}; Hathaway et al., \cite{hathaway}) because, if these cells exist, they should be present in deeper layers. The ratio of inverse Rossby number, i.e. the Coriolis number, of giant cells and supergranules, was estimated by Komm et al. (\cite{komm}) to be about 60. Komm et al. (\cite{komm}) found a similar value for the covariances of sunspots and small magnetic features, and these differences were also reported by Meunier et al. (\cite{meunier}). Komm et al. (\cite{komm}) interpreted these differences by assuming that the sunspot magnetic fields were anchored in the deep giant cells whereas the small magnetic fields were only influenced by the near-surface supergranules and the ratio of Reynolds stresses in these two regions was found to be close to 60. If this interpretation is correct, then the assumed impact of the tachoclyne-zone shear oscillation on the giant cells, and indirectly on the sunspot motion correlations, may be studied by selecting the periods of opposite phases in the shear oscillations; this can only be attempted, however, in a future study by having a reasonably long overlap between the DPD catalogue and shear oscillation data.

\begin{acknowledgements}
      This work was supported by the ESA PECS project No. 98017. Thanks are due to L. Gy\H ori and T. Baranyi for their enormous efforts to produce the DPD catalogue.
\end{acknowledgements}

\end{document}